\documentclass[conference]{IEEEtran}

\pdfoutput=1
\usepackage[bookmarks=false]{hyperref}

\usepackage{cite}
\usepackage{amsmath,amssymb,amsfonts}
\usepackage{algorithmic}
\usepackage{graphicx}
\usepackage{textcomp}
\usepackage{xcolor}
\usepackage{multicol}
\usepackage{flushend}

\usepackage{todonotes}

\def\BibTeX{{\rm B\kern-.05em{\sc i\kern-.025em b}\kern-.08em
    T\kern-.1667em\lower.7ex\hbox{E}\kern-.125emX}}

\begin{document}

\title{Intrusion Detection and Localization for Networked Embedded Control Systems}

\author{\IEEEauthorblockN{Vuk Lesi,
Marcio Juliato, Shabbir Ahmed, Christopher Gutierrez, Qian Wang, Manoj Sastry}
\IEEEauthorblockA{Intel Corporation\\
Hillsboro, OR, USA \\
\{vuk.lesi, marcio.juliato, shabbir.ahmed, christopher.n.gutierrez, qian4.wang, manoj.r.sastry\}@intel.com}}


\maketitle

\begin{abstract}

Closed-loop control systems employ continuous sensing and actuation to maintain controlled variables within preset bounds and achieve the desired system output. Intentional disturbances in the system, such as in the case of cyberattacks, can compromise reachability of control goals, and in several cases jeopardize safety. The increasing connectivity and exposure of networked control to external networks has enabled attackers to compromise these systems by exploiting security vulnerabilities. Attacks against safety-critical control loops can not only drive the system over a trajectory different from the desired, but also cause fatal consequences to humans.
In this paper we present a physics-based Intrusion Detection System (IDS) aimed at increasing the security in control systems. In addition to conventional process state estimation for intrusion detection, since the controller cannot be trusted, we introduce a controller state estimator. Additionally, we make our detector context-aware by utilizing sensor measurements from other control loops, which allows to distinguish and characterize disturbances from attacks. We introduce adaptive thresholding and adaptive filtering as means to achieve context-awareness. Together, these methodologies allow detection and localization of attacks in closed-loop controls. Finally, we demonstrate feasibility of the approach by mounting a series of attacks against a networked Direct Current (DC) motor closed-loop speed control deployed on an ECU testbed, as well as on a simulated automated lane keeping system. Among other application domains, this set of approaches is key to support security in automotive systems, and ultimately increase road and passenger safety.
\end{abstract}

\begin{IEEEkeywords}
Autonomous Systems Security; Automotive Security; Intrusion Detection Systems; Closed-Loop Control; Networked Control Systems Security
\end{IEEEkeywords}

\section{Introduction}
\label{sec:intro}

\begin{figure}[!t]
    \begin{center}
    \includegraphics[width=0.48\textwidth]{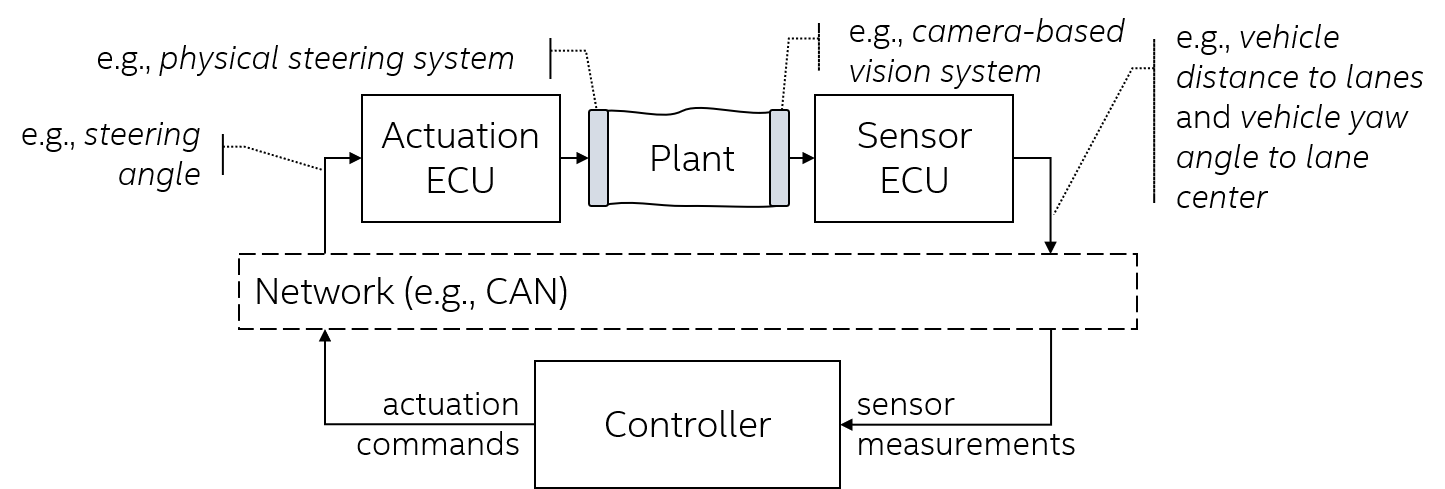}
    \caption{Standard networked control architecture.}
    \label{fig:controlLoop}
    \end{center}
\end{figure}

The evolution of embedded networked control systems has enabled significant design changes in autonomous systems. For instance, closed-loop controls along with In-Vehicle Networks (IVNs), as illustrated in Figure~\ref{fig:controlLoop}, are the underlying enablers of modern longitudinal and lateral controls in Advanced Driver-Assistance Systems (ADAS) and Automated Driving Systems (ADS). The ability to sense physical processes with modern high-resolution sensing devices and in feedback affect the same processes based on control laws constitutes the foundation of all Cyber-Physical Systems (CPS).
However, the inherent physical interaction with the environment poses severe safety risks when exposed to cyber threats. Unlike information systems where attacks typically lead to data or financial loss, attacks on CPS have a physical manifestation on autonomous systems and directly impact safety.

The widespread use of closed-loop control began in industrial applications, where components of the electrical grid, chemical process control and manufacturing systems were automated to increase productivity. These processes warranted significant research attention due to their sensitivity to downtime on one hand, and potential disastrous consequences in the case of control failure on the other. While usage of security measures is reported in industrial control applications~\cite{FDI0}, we witnessed cyberattacks that demonstrated the fragility of automotive systems ~\cite{attacksSurvey,jeepHack,teslaHack}. In modern vehicles, control networks are not isolated; multiple control networks (e.g., powertrain, chassis) are typically interconnected through a gateway, which may also be connected to modules (e.g., In-Vehicle Infotainment (IVI)) that enables connectivity to external networks such as cellular, \mbox{Wi-Fi}, and Bluetooth. The compromise of interconnecting nodes has shown to be a common attack vector in attacks on automotive systems. Once an attacker has gained foothold in one of those modules, malicious messages can be transmitted on safety-critical control networks. The feasibility of such attacks~\cite{attacksSurvey} has shown that traditional security practices are mandatory, but not sufficient, to comprehensively secure safety-critical subsystems. Moreover, networked control systems offer attackers the opportunity not only to disable the system (e.g., causing denial of service by flooding the control network), but to divert the system over a trajectory different than desired (e.g., by spoofing sensor measurements and actuation commands). For instance, by spoofing speed sensing measurements, a cruise controller can be led into a cruise speed inconsistent with the context of the road and speed limits.

The knowledge of the closed-loop control system offers unique opportunities to efficiently protect these systems. The history of system’s states along with the control input that is currently being applied defines the output of the system, which should reflect the expected system dynamics. For instance, if the current vehicle speed is $50\:mph$, and a certain throttle position is commanded, vehicle speed in the next moment cannot be $70\:mph$ (even the measurement noise levels would be unrealistic for this to happen). Namely, physics-based IDS utilize the system’s dynamical behavior to cross-check expectations with live sensing data by performing state estimation. The difference between the predicted (i.e., based on the model) and the observed output (i.e., sensor measurements) forms a residual signal that is used as an indicator of anomalous behavior. However, while this core principle is applied in state-of-the art anomaly detection (e.g., both for failure~\cite{FDI1,FDI2,FDI3} and attack detection~\cite{chi2application1,chi2application2,chi2application3,chi2application4,jovanov2017sporadic}), the common assumption in these works is that the attacker is present within the sensor’s platform (i.e., the networked Electronic Control Unit (ECU) executing the sensing task). From that standpoint, the attacker is able to alter sensor measurements before they are sent over the network to the controller (i.e., the networked ECU executing the control task). While this model of an internal sensor attacker is valid, it is not comprehensive; i.e., the attacker performing false data injection may also reside in a different ECU, without the capability to disable the original sensor. The same holds true for actuation.

Concurrently, it is commonly assumed that the controller is trusted, i.e., that an IDS can be deployed alongside the controller implementation (and thus have direct, trusted access to controller's state). This assumption is often invalid in practice, where legacy systems were not designed to properly satisfy security requirements, and where common vulnerabilities simultaneously manifest in multiple  ECUs (e.g., a vulnerability in the underlying operating system). Consequently, in this paper we consider the problem of detecting and localizing false data injection attacks that can be both \emph{internal} and \emph{external} to the sensing/control/actuation ECUs.

Our approach utilizes controller state estimation in addition to exclusive plant\footnote{We will use the standard terminology to refer to the physical process under control as the plant.} state estimation in prior work. Firstly, we observe the controller (and not only the plant) as a dynamical system for purposes of intrusion detection. Additionally, we propose a layer of context-awareness to our detector to minimize false positives in scenarios where operating conditions or failures may appear similar to attacks. These extensions to the design of the physics-based intrusion detector allows us to: (1) Perform attack detection and localization by patterning residual signals not only characterizing plant, but also controller deviation from normal behavior; (2) Not fully rely on the security of the controller ECU; and (3) Provide a fully centralized (i.e., network-attached) intrusion detection solution (including localization) that utilizes physical models.

This paper is organized as follows. In Section~\ref{sec:motivation} we define the problem considered by this work. Section~\ref{sec:overview} gives a short introduction to closed-loop control before detailing our approach to physics-based intrusion detection in closed-loop control systems in Section~\ref{sec:IDS}. An experimental case study is presented in Section~\ref{sec:caseStudy} before related work (Section~\ref{sec:relatedWork}), while Section~\ref{sec:conclusion} summarizes our conclusions.

\section{Motivation and Problem Statement}
\label{sec:motivation}

In this section, we start by outlining the gaps in state-of-the-art physics-based intrusion detection, before defining the adversary model and stating the specific problems we tackle in this work. Physics-based approaches for monitoring operation of closed-loop control systems were originally developed for purposes of implementing Failure Detection and Isolation (FDI)~\cite{FDI1,FDI2}. The basic principle underlying physics-based approaches is analytical redundancy. Analytical redundancy leverages the possession of physical models of the system that are required during the control design phase by executing these models to check if currently observed output of the system, as reported by the sensors, corresponds to physically attainable values computationally obtained from the model. Analytical redundancy is typically inexpensive, since execution of dynamical models is based on relatively simple mathematical operations. For example, lateral vehicle dynamics around a given speed can be described by a fourth-order linear dynamical model~\cite{vehicleControl}. In this case, calculating the output of the model reduces to performing several multiplications and additions, while keeping track of the most recent computed outputs. The difference between the predicted and the modeled output forms a residual signal; true equality among predicted and observed outputs cannot be expected even for a fully functional system due to noise and modeling imperfections (i.e., residual signals are never constantly zero). Therefore, their statistical properties (e.g., running mean, power) are monitored through hypothesis testing.

This detection mechanism appears both in FDI publications~\cite{FDI1,FDI2,FDI3} and in security-related work~\cite{chi2application1,chi2application2,chi2application3,chi2application4,jovanov2017sporadic}. However, prior work assumes that the controller is secure, and a decentralized, controller-centric implementation of the aforementioned detector is utilized without controller state estimation. This assumption can be easily violated given the hostile environment in which vehicular networked control systems operate (e.g., exposure to external networks, potential common vulnerabilities~\cite{attacksSurvey}). Moreover, state-of-art physics-based anomaly detection does not take into account network-based anomalies (attacks) where additional malicious control commands or sensor measurements are injected into the control system. Furthermore, a distributed anomaly detection scheme is assumed that requires controller-centric implementation (i.e., each and every controller ECU implements intrusion detection for the control loops it controls). In this case, a software-based attacks on controllers can also compromise the intrusion detectors for those respective control loops. The following subsection specifies the considered attack model which relaxes the unrealistic assumptions.

\subsection{Attack Model}
\label{subsec:attackModel}

We assume that all components of the closed-loop can be attacked, i.e., all ECUs (sensor, controller, actuator), as well as the physical sensing/actuation components tied to the plant. Furthermore, we do not limit attacker’s timing or accessibility to sensing and/or actuation signals; the attacker may alter or inject any or all sensor measurements sent by sensor ECUs (used by controllers), or any or all actuation commands sent by controllers (used by actuators) at any time. In the following, we detail the attack model from the entry point perspective.

\emph{Internal software attack} is performed by an attacker who has gained software execution privileges on an ECU of the system (e.g. by exploiting a security vulnerability). We differentiate three types of attacks based on which component has been compromised. In the case of the sensor node, the attacker alters (or blocks) sensor measurements before they are transmitted to the controller. This type of attack is also known as \emph{message modification}. In case the target ECU is the controller, the attacker tampers with the control algorithm. If the actuator node is attacked, the attacker can alter control commands before they are applied on the corresponding physical actuator.

\emph{External software attack} is performed by an attacker who is able to execute software on an ECU that interfaces the control bus in order to impersonate other ECUs participating in closed-loop control. The vast majority of modern vehicles use Controller Area Network (CAN) bus to interconnect safety-critical control components. CAN is a broadcast-based network communication protocol that does not intrinsically support message authentication; consequently, any network node can transmit messages on behalf of another nodes. This type of attack is also known as \emph{masquerading}. The actuator ECU interfaces directly with the physical process and it usually does not send messages to the bus, and therefore this kind of attack does not apply on actuator ECUs. In the case of sensor impersonation, the attacker sends malicious sensor messages alongside original measurements. If the target node is the controller, the attacker sends malicious actuation commands alongside original ones. Attacks where the adversary is able to physically connect to the network (e.g., via a diagnostic port) are also captured by this class of attacks.


\begin{figure}[!t]
    \begin{center}
    \includegraphics[width=0.48\textwidth]{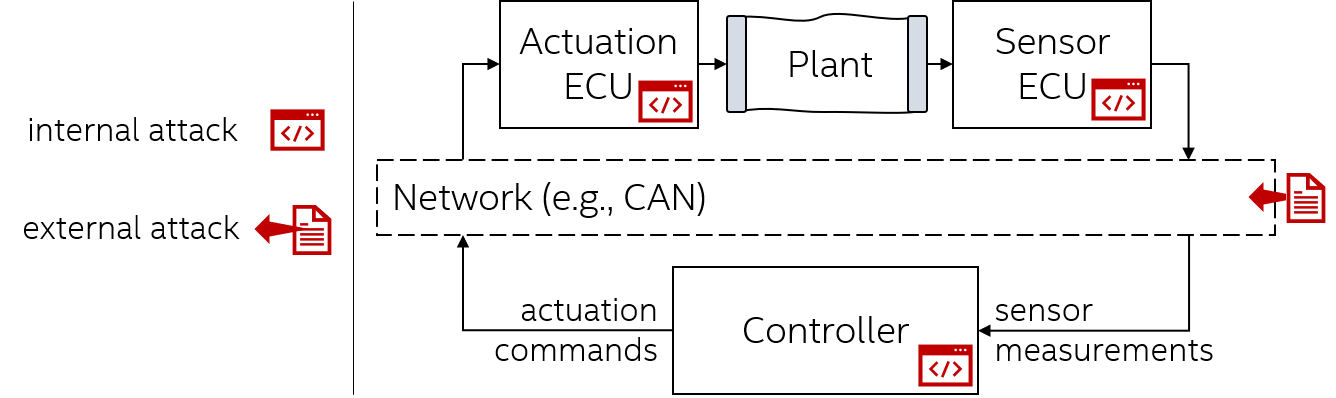}
    \caption{Attack entry points on the standard networked control architecture.}
    \label{fig:attackModel}
    \end{center}
\end{figure}

In summary, from the perspective of a closed-loop control system, the malicious interference can be characterized by
\paragraph*{Target}
\begin{itemize}
    \setlength{\itemindent}{2em}
    \item Sensor,
    \item Controller,
    \item Actuator.
\end{itemize}
\paragraph*{Class}
\begin{itemize}
    \setlength{\itemindent}{2em}
    \item Internal attacker (software execution-based),
    \item External attacker (network-based injection).
\end{itemize}

Sensor measurements and actuation commands can be tampered with at the physical process level; e.g., through electromagnetic fields, mechanical forces, temperature. These attacks reflect on the operation of the closed-loop similarly to internal attacks on sensor and/or actuator ECUs.

\subsection{Problem Statement}

In order to enable detection and localization of attacks specified in Section~\ref{subsec:attackModel} in realistic deployment scenarios, the legacy controller ECU cannot be trusted. Additionally, the employed physical models may be less than optimal when the plant is subject to external disturbances from the environment (e.g., unwanted forces on the vehicle resulting from side wind or road inclination), which (as described in Section~\ref{subsec:contextAwareness}) would cause false alarms. Consequently, the following challenges arise:

\begin{itemize}
    \item How can the physics-based IDS provide not only attack detection but also localization; i.e., how to distinguish different types of attacks against ECUs in the closed-loop control network?
    \item How can the intrusion detection and localization system be made \emph{driving context-aware}, in order to minimize false alarms?
\end{itemize}

In the following sections, we describe our solutions to challenges above by introducing to a centralized, network-based implementation of a context-aware detector based on residual patterning. Notice that, while we use examples from the automotive domain to motivate and develop our presentation, the attack model as well as the intrusion detection and localization solution are applicable in other domains (e.g., industrial, IoT, robotics).

\section{Overview of Closed-Loop Control}
\label{sec:overview}
Closed-loop control systems (or feedback control systems) are designed to actively compensate environmental and process conditions in order to maintain the desired system’s trajectory. This is typically achieved with three distinct components: sensors, actuators, and a controller. Sensors provide measurements of controlled variables while the controller consumes this information to determine how well the current output of the plant corresponds to the desired one. As a result, the controller drives actuators that correspondingly affect the physical process. Through the dynamics of the process, these actions are reflected on sensor measurements when the control loop is executed again. Periodical execution of this sequence of (1) performing measurements on the physical process, (2) executing the control algorithm, and (3) actuating provides the desired control over the physical variables of interest. 

The functional purpose of the control system is to keep the observed (or estimated) system output as close to the desired as possible, sometimes with specific requirements for transient responses. Properties of the response to controller commands depend on the dynamics of the system and the controller; this implies that special care is required during controller design. For example, it is not desired for a lane keeping system to have overshoot in lateral positioning since this would the vehicle to move around the lane potentially impacting comfort and safety. Figure~\ref{fig:controlLoop} depicts the standard networked control architecture where sensing, actuation and control nodes communicate over a shared network. To provide a running example in this paper, the figure also maps the generic components to a specific safety-critical ADAS system --- lane keeping, where CAN networks are used.

\begin{figure}[!t]
    \begin{center}
    \includegraphics[width=0.48\textwidth]{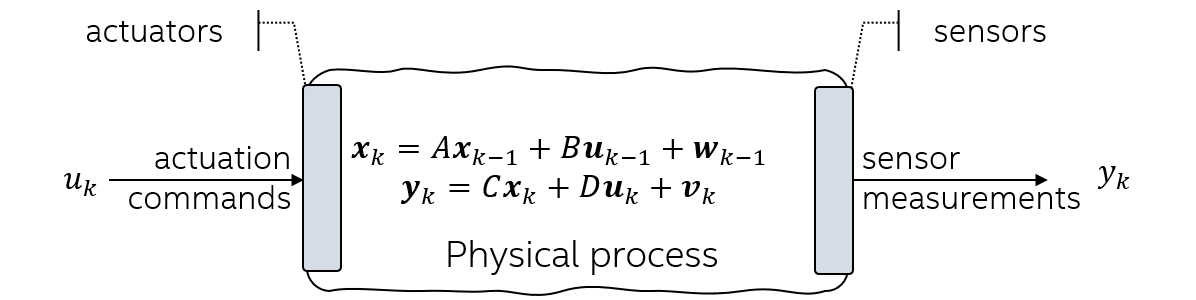}
    \caption{Linear, Time-Invariant (LTI) plant model.}
    \label{fig:LTIplant}
    \end{center}
\end{figure}

All plants have some dynamical behavior intrinsic to their first principles of operation. For instance, no matter how abrupt the steering is applied, the vehicle's angle to the center of a highway lane cannot change by a large angle instantly. Behavior of dynamical systems can be captured by specifying the plant's state of interest (e.g., vehicle's lateral position relative to the lane) dependency on the previous state and the previous control commands (e.g., steering angle). The plant's states are variables that, when combined, fully describe system’s response to inputs. In the case of lateral control for automated lane keeping, the set of physical states sufficiently describing lateral motion of an vehicle are \emph{lateral positioning error} (i.e., distance from vehicle's center to center of the lane), the \emph{yaw angle error} (i.e., angular distance from vehicle's heading to lane direction), and their \emph{derivatives} (i.e., rates of change). Figure~\ref{fig:LTIplant} shows the mathematical representation of the stochastic (i.e., featuring random noise terms on the plant's states and sensor measurements),  discrete-time Linear Time-Invariant (LTI) state-space model of a plant.\footnote{For improved clarity, we limit our presentation to Linear Time-Invariant (LTI) systems. While LTI models can be used to capture a large class of systems’ dynamical behavior, our approach (detailed in Section~\ref{sec:IDS}) is not fundamentally limited to linear dynamics.}

In this case, system behavior is modeled as:
\begin{align}
 x_{k+1}=& Ax_{k}+Bu_{k}+w_{k},\\
 y_k=& Cx_k+Du_k+v_k,   
 \label{eq:LTI}
\end{align}
where $x_k$ is the plant state vector (dimension $n\times 1$), $A$ is the state transition matrix (dim. $n\times n$), $u_k$ is the vector of system inputs (dim. $m\times 1$), $B$ is the input matrix (dim. $n\times m$), $y_k$ is the vector of sensor measurements (dim. $p\times 1$), $C$ is the measurement matrix (dim. $p\times n$), and $D$ is the matrix directly mapping inputs to outputs (dim. $p\times r$). Here, $k$ indicates the time instant as a integer multiple of the sampling period of the system $T_s$. The sampling period of the system is dependent on its dynamics and denotes the time period at which the physical process is sampled through sensors and affected through actuators.


This type of model is widely used in control-theoretic research for analysis of dynamical systems. Intuitively, state matrix $A$ encodes the dependency of the current state of the system on the most recent previous state. Matrix $B$ maps the effects of available actuators (plant inputs) on the states of the system. Matrix $C$ maps outputs of the plant into measurable values while matrix $D$ maps actuation inputs into states. $w_k$ is the process noise which captures all uncertainty in the plant model while $v_k$ is the measurement noise capturing uncertainty of sensors. They are often assumed to be independent and normally distributed. Such a model can be obtained from first principles, i.e., from physical analysis of the system~\cite{controlTutorials}, or from system identification~\cite{systemIdentification}.

In the case of our running example, automated lane keeping, a passenger vehicle can be modeled as~\cite{vehicleControl}
\begin{align}
\label{eq:laneKeeping}
    A&=\begin{bmatrix}0 & 1 & 0 & 0\\
                     0 & -\frac{2c_f+2c_r}{mV_x} & \frac{2c_f+2c_r}{m} & \frac{-2c_fl_f+2c_rl_r}{mV_x}\\
                     0 & 0 & 0 & 1\\
                     0 & - \frac{2c_fl_f-2c_rl_r}{I_zV_x} & \frac{2c_fl_f-2c_rl_r}{I_z} &  -\frac{2c_fl_f^2+2c_rl_r^2}{I_zV_x}
        \end{bmatrix}\\
    B& = \begin{bmatrix}
                0\\
                2\frac{c_f}{m}\\
                0\\
                \frac{2c_fl_f}{I_z}
        \end{bmatrix},
\end{align}
where $m\:[kg]$ is the total mass of vehicle, $l_f\:[m]$ is the longitudinal distance from center of gravity to front tires, $l_r\:[m]$ is the longitudinal distance from center of gravity to rear tires [m], $L_w = l_f+l_r\:[m]$ is the vehicle Wheelbase, $m_r = m\frac{l_f}{L_w}\:[kg]$ and $m_f = m\frac{l_r}{l_w}\:[kg]$ are masses carried on the rear and front axles, respectively, $c_f,\:c_r\:[\frac{N}{rad}]$ is the cornering stiffness of front and rear tires, $I_z\:[kg\cdot m^2]$ is the yaw moment of inertia. Since the model of the system is intrinsically nonlinear, the linear representation in~\eqref{eq:laneKeeping} holds around a fixed cruising speed of $V_x\:[\frac{m}{s}]$. Here, the control signal is the steering angle while a camera-based sensing system provides measurements of the distance to center of the lane and yaw angle to center of the lane. Therefore, the measurement matrix $C$ is
\begin{equation}
\label{eq:endLaneKeeping}
    C=\begin{bmatrix}1 & 0 & 0 & 0\\
                     0 & 0 & 1 & 0\\
        \end{bmatrix}.
\end{equation}
State noise can be extracted from certainty in vehicle parameters, while sensing noise can be extracted from the sensor's documentation.

Notice that this model is derived from first principles, i.e., from deep knowledge of the physics of the underlying process (as detailed in~\cite{vehicleControl}). On the other hand, system identification (e.g.,~\cite{systemIdentification}) can be used to determine the physical models, if the detailed structure or tuning of the plant are not known. We use this principle to demonstrate practical feasibility in our case study (Section~\ref{sec:caseStudy}).

\subsection{State Estimation}
\label{subsec:kalman}
Since the initial state of the system may not always be known, and since not all states of interest can always be measured, state estimation techniques are applied. Kalman filter~\cite{kalman} is one of the most broadly used estimators and is optimal for plants as in~\eqref{eq:LTI} since it minimizes the mean square estimation error. Kalman filter can estimate system’s states based on the model at hand and the current output of the system (obtained from sensors). For purposes of illustrating the state estimation principle, we will omit process and measurement noises $w_k$ and $v_k$ from equations that follow. Kalman filter estimates plant’s state in two steps. In the first step the state of the system is \emph{predicted} based on the model matrices from~\eqref{eq:LTI}, as the current state of the model, given the previous estimation
\begin{equation}
    \hat{x}_k^{predicted}=A\hat{x}_{k-1}+Bu_{k-1},
\end{equation}
where $\hat{x}_k$ denotes the current state estimated by the filter. The estimation error covariance $P$ is predicted as
\begin{equation}
    P_k^{predicted}=AP_{k-1}A^\top+Q,
\end{equation}
where $Q$ is the covariance matrix of the process noise $w_k$. Then, a new measurement sample from the sensors is used to \emph{update} the prediction in three sub-steps. Firstly, the Kalman gain is computed
\begin{equation}
    K_k=P_k^{predicted}C^{\top} (CP_k^{predicted} C^{\top}+R)^{-1},    
\end{equation}
where $R$ is the covariance matrix of the measurement noise $v_k$. Then, the state estimate is corrected with the obtained measurement $y_k$
\begin{equation}
\hat{x}_k=\hat{x}_k^{predicted}+K_k (y_k-C\hat{x}_k^{predicted}).
\label{eq:stateEstimate}
\end{equation}
Finally, the state estimation error covariance matrix is updated as
\begin{equation}
    P_k=(I-K_k C)P_k^{predicted}.
\end{equation}
The Kalman gain matrix $K_k$ converges quickly and remains constant for systems described with~\eqref{eq:LTI}. Notice that an initial estimate of the state $\hat{x}_{-1}$ and the state estimation error covariance $P_{-1}$ is needed. If the initial state of the system is known, that known state should be assigned to $\hat{x}_{-1}$, and $P_{-1}$ may be kept low according to the certainty into the initial state estimate. On the other hand, if the initial state is not at all known, an empirical approach has been to initialize $P_{-1}$ to a very large number; this models high uncertainty in the initial state. Initial version of the covariance matrix $P_{-1}$ can be tuned experimentally.\footnote{Extended and Unscented Kalman Filter (EKF/UKF) can be used for non-linear state estimation.}

The essential parameter used for intrusion detection is the residual signal, i.e., the difference between the predicted output of the system and the measured (or reported) output as used in~\eqref{eq:stateEstimate}
\begin{equation}
    r_k=y_k-C\hat{x}_k^{predicted}.
\end{equation}

The residual is a zero-mean normally distributed random variable with a constant covariance matrix \mbox{$\Sigma=CP_k^{predicted} C^\top+R$} under no-attack conditions, which facilitates attack detection since statistical properties of $r$ change under attack conditions. The power of this signal is convenient to observe because it is $\chi^2$ distributed with $m$ degrees of freedom (recall that $m$ is the number of sensors in the system). Intrusion detection can be formulated as a hypothesis testing problem where the residual power is tested to be below or above a predefined threshold $g$
\begin{equation}
\label{eq:residThreshold}
r_k^\top \sigma^{-1} r_k\gtrless g.    
\end{equation}
For instance, $g$ can be set so that a desired (low) false alarm rate is exhibited by the detector (since the left side of the hypothesis testing inequality is a random variable of known distribution). This can be achieved by computing the inverse cumulative density function (of the $\chi^2$ distribution) at $1-P_{false\:alarm}$, where $P_{false\:alarm}$ is the desired false alarm probability (e.g., $10^{-6}$).
\begin{equation}
    g = cdf^{-1}(1-P_{false\:alarm})
\end{equation}
Notice also that $g$ is a vector of thresholds that can also be experimentally set for each sensor such that desired performance is achieved. In Section~\ref{subsec:contextAwareness} we discuss how the threshold can be adapted to account for the effect of disturbances on residuals. Besides the $\chi^2$ detector, other detection approaches are also proposed~\cite{chi2application1,chi2application2,chi2application3,chi2application4} on top of the same basic estimation principle. 


\section{Physics-Based Anomaly Detection and Localization}
\label{sec:IDS}

\begin{figure}[!ht]
    \begin{center}
    \includegraphics[width=0.46\textwidth]{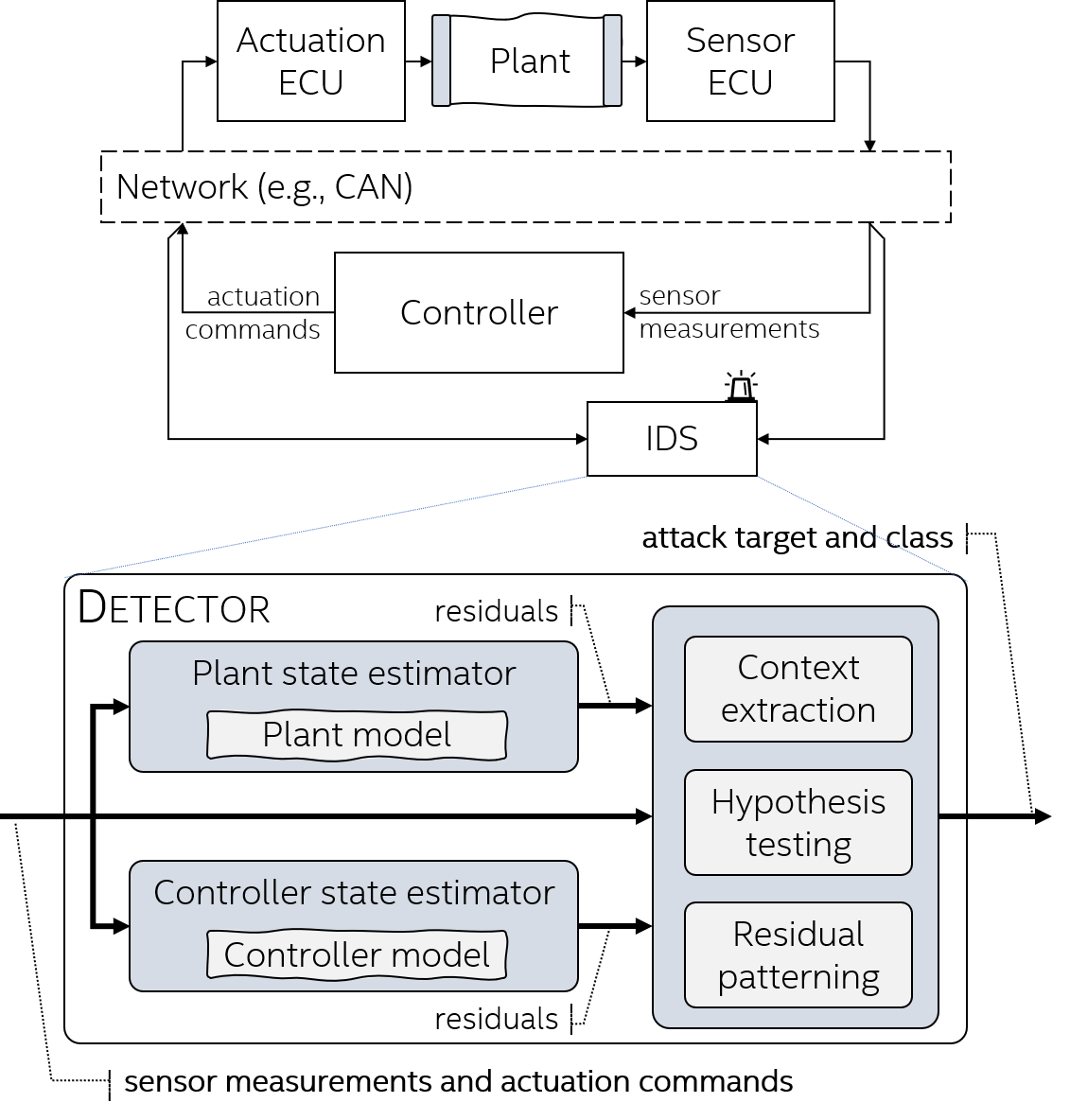}
    \caption{Block diagram of our solution to physics-based intrusion detection and localization.}
    \label{fig:IDSarchitecture}
    \end{center}
\end{figure}

Before detailing our solution, we provide a brief summary, illustrated as a block diagram in Figure~\ref{fig:IDSarchitecture}. Sensor measurements and actuation commands transmitted over the control network are utilized to perform plant state estimation (presented in Section~\ref{subsec:kalman}) and controller state estimation (to be discussed in Sec~\ref{subsec:controllerEstimation}). Additional sensor measurements are utilized to extract indicators of operational context (i.e., to aid distinguishing disturbances from attacks, described in Section~\ref{subsec:contextAwareness}), and together with residuals generated during state estimation, thresholded with corresponding values. Finally, the pattern of affected signals (i.e., ones exceeding thresholds) is matched to a specific pattern (in residual patterning), to unveil the target and class of the attack. Notice that, in terms of deployment, the entire detector can be implemented in software, and, as described earlier, its architecture allows independence from any existing ECUs part of the protected closed-loop.

The key ingredient to physics-based anomaly detection is the state estimation process leading to residual generation, as described in Sec~\ref{subsec:kalman}. For our running example (i.e., automated lane keeping system described in 
\eqref{eq:laneKeeping}-\eqref{eq:endLaneKeeping}), we developed a Matlab-Simulink-based simulation employing Linear Quadratic Gaussian (LQG) optimal control~\cite{linearOptimalControlBook}, as well as a Kalman-based residual generator for detecting attacks (as described in Section~\ref{subsec:kalman}). Recall that plant's sensors (e.g., camera-based vision) are providing \emph{lateral error} and \emph{yaw angle error} to the controller (i.e., controller's goal is to maintain vehicle as close to the center of the lane as possible), and the controller computes the required steering angle. Within this simulation environment, we are able to model all attacks enumerated in the attack model (Section~\ref{subsec:attackModel}). We begin by pointing out the gap in state-of-the-art physics-based intrusion detection.

\begin{figure}[!hb]
    \begin{center}
    \includegraphics[width=0.49\textwidth]{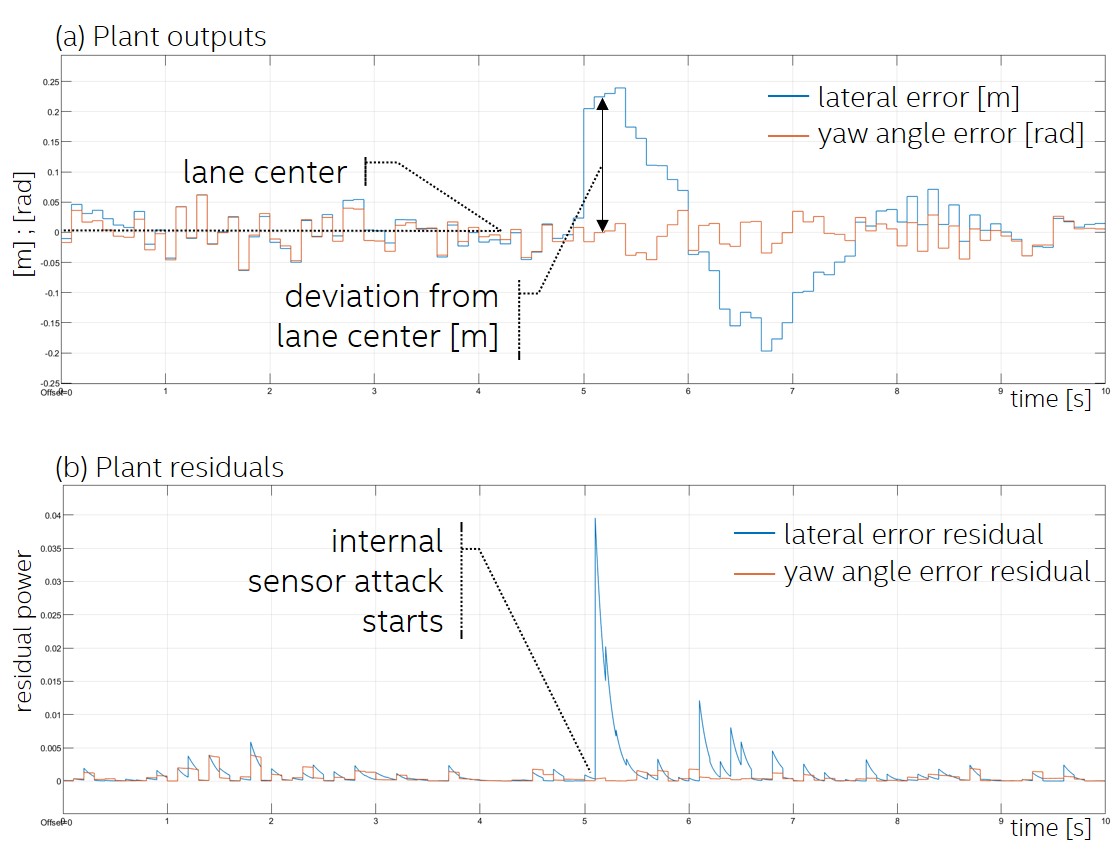}
    \caption{Internal sensor attack on automated lane keeping: after $t=5s$, attacker biases lateral position measurement by $0.2m$.}
    \label{fig:internalSensorAttack}
    \end{center}
\end{figure}

Figure~\ref{fig:internalSensorAttack}(a) shows that an internal attack on the sensor ECU successfully propagates to system's outputs, shifting the vehicle towards the left edge of the lane by $0.2m$ after $t=5s$. On the other hand, Figure~\ref{fig:internalSensorAttack}(b) shows that this behavior can be successfully flagged as malicious, based on the plant residuals. Intuitively, by biasing sensor measurements, the attacker \emph{offsets} plant's behavior, which, due to this attack, does not conform to the physical model.

\begin{figure}[!t]
    \begin{center}
    \includegraphics[width=0.49\textwidth]{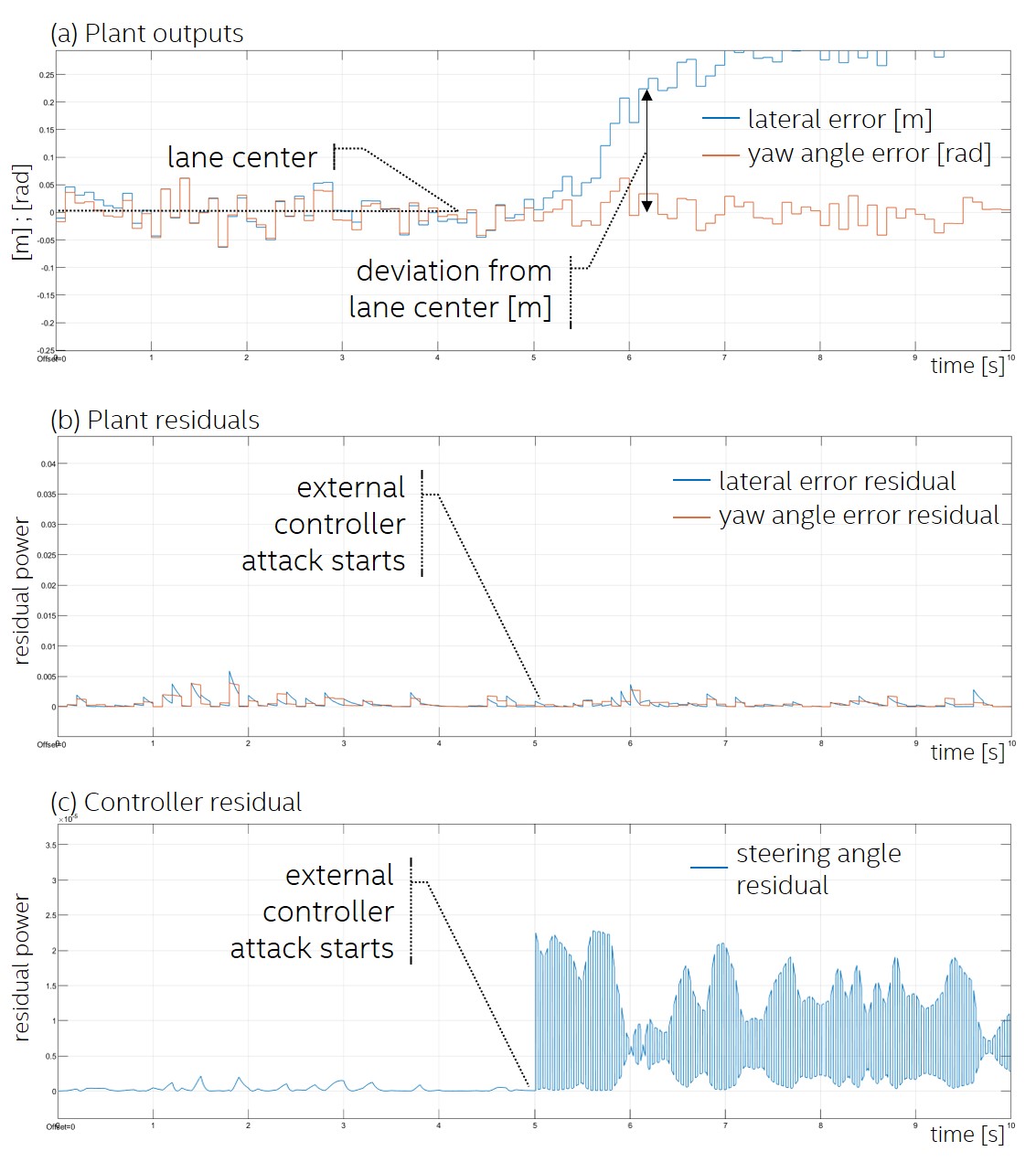}
    \caption{External controller attack on automated lane keeping: after $t=5s$, attacker injects steering commands to shift the vehicle towards the left edge of the lane.}
    \label{fig:injectionActuationAttack}
    \end{center}
\end{figure}

However, plant state estimation is not sufficient to provide high attack diagnostic coverage (i.e., some attacks cannot be detected), and consequently lacks attack localization capabilities. Figure~\ref{fig:injectionActuationAttack}(b) shows that plant residual signals are unaffected by the external controller attack, i.e., when the attacker injects additional actuation commands into the network, while the plant's outputs are heavily affected and the vehicle departs its lane (shown in Figure~\ref{fig:injectionActuationAttack}(a)). In this case, the attacker injects messages at $10\times$ the frequency of the original control loop sampling period (i.e., periodicity of the actuation message), which therefore causes the actuator ECU to consume both actuation commands from the legitimate controller ECU, as well as attacker's.\footnote{In this simulation, we assumed that the actuator ECU will reject actuation commands arriving more frequently than $10\times$ the original sampling period ($100ms$), however, a different platform model can easily be incorporated.} Intuitively, the plant residuals are not affected as the plant is not malfunctioning---it is legitimately reacting according to the underlying physical laws but in this case to both the commands of the legitimate controller and the attacker. However, the controller is attacked, and therefore it is critical to monitor the controller's state estimate in order to detect attacks such as message modification or masquerading on behalf of the controller. We now introduce our approach to detect these attacks.

\subsection{Controller State Estimation}
\label{subsec:controllerEstimation}
Controllers can also be described with~\eqref{eq:LTI}, with two modeling subtleties. Firstly, matrix $D$ in model~\eqref{eq:LTI} is typically non-zero for controllers since it models proportional gain of the controller. Secondly, if the controller implementation is known (i.e., it is not determined through system identification) "process" noise can be considered zero due to high certainty into evolution of controller's states.

In the case of our running example, Figure~\ref{fig:injectionActuationAttack}(c) shows that the controller residual provides a clear indication of the attack. Ultimately, we threshold residual signals and match the combination of binary detection flags after thresholding to not only detect, but localize the attacker. Due to space constraints, a detailed attack characterization map is given for our case study (Section~\ref{sec:caseStudy}).

\subsection{Context-Aware Detection}
\label{subsec:contextAwareness}
Disturbances encompass all effects on the output of the plant that operating conditions may impose. Controllers are typically designed to suppress a certain type of disturbance; e.g., a cruise controller is designed to suppress effects of increasing road inclinations (by increasing the throttle opening), and a lane keeping controller is designed to suppress effects of side winds and banking (by compensating the steering angle). In these cases, the vehicle's sensor residual signals would be similarly affected as in the scenario in which an attacker is biasing sensor measurements. Therefore, the IDS must be context-aware in order to distinguish attacks from non-malicious disturbances, which can be achieved by employing additional available sensing streams. Intuitively, the more (relevant) sensing streams are available, the richer context reconstruction is possible, increasing detection accuracy.

\subsubsection{Adaptive Thresholding}
\emph{Adaptive thresholding} leverages context information available from other sensors to adjust thresholds for residual signals to current operating conditions. The threshold can be made arbitrarily dependent on the value of another sensor, formally, the predefined threshold in~\eqref{eq:residThreshold} can be generalized
\begin{equation}
    g = f_{thr}(s1, s2, ...),
\end{equation}
where $s1$, $s2$, ... are additional sensor measurements not directly used by the monitored control loop, and $f_{thr}$ is an arbitrary function. For instance, in the case of our running example, the automated lane keeping control loop is requesting a specific steering angle from the electrically-assisted steering subsystem which can return the steering motor torque required to achieve the required angle. This torque will depend on the operating condition, which, in the absence of attack, will correspond to road conditions. On the other hand, if the plant residuals are indicating an anomaly, and the torque required to complete current actions is unaffected (i.e., there is no resistance from road conditions), then an attack can be flagged. Finally, if attack conditions persist together with disturbances, the design of the adaptive threshold (i.e., its dependence on valid operating conditions) ensures the corresponding residual signal will exceed it. Intuitively, threshold granularity (time and accuracy) is closely related to the granularity of the additional sensors used to adjust the threshold.

\subsubsection{Adaptive Estimation}
\emph{Adaptive estimation} leverages knowledge of disturbance models---these models capture dependence of the system model on operating conditions. For instance, matrices of the vehicle model from~\eqref{eq:laneKeeping} can be adapted  output disturbances. Thus, the model can be adjusted to account for the current operating condition. For our running example, a physical relationship between the torque required to complete the steering command and the resulting vehicle position (i.e., lateral position and yaw angle relative to lane) is needed. Formally, plant matrices as used in~\eqref{eq:LTI} can be generalized
\begin{align*}
    A = f_A(s1, s2, ...), B = f_B(s1, s2, ...),\\
    C = f_C(s1, s2, ...), D = f_D(s1, s2, ...),
\end{align*}
where $s1$, $s2$, ... are additional sensor measurements not directly used by the monitored control loop, and $f_{A}$,$f_{B}$,$f_{C}$,$f_{D}$ are arbitrary functions. Intuitively, by adding more context information in the model, the plant state can be accurately estimated in a broader range of conditions. Due to space constraints, we illustrate how both context-awareness techniques can be utilized on our case study in Section~\ref{sec:caseStudy}.



\section{Case Study}
\label{sec:caseStudy}

\begin{figure}[!ht]
    \begin{center}
    \includegraphics[width=0.48\textwidth]{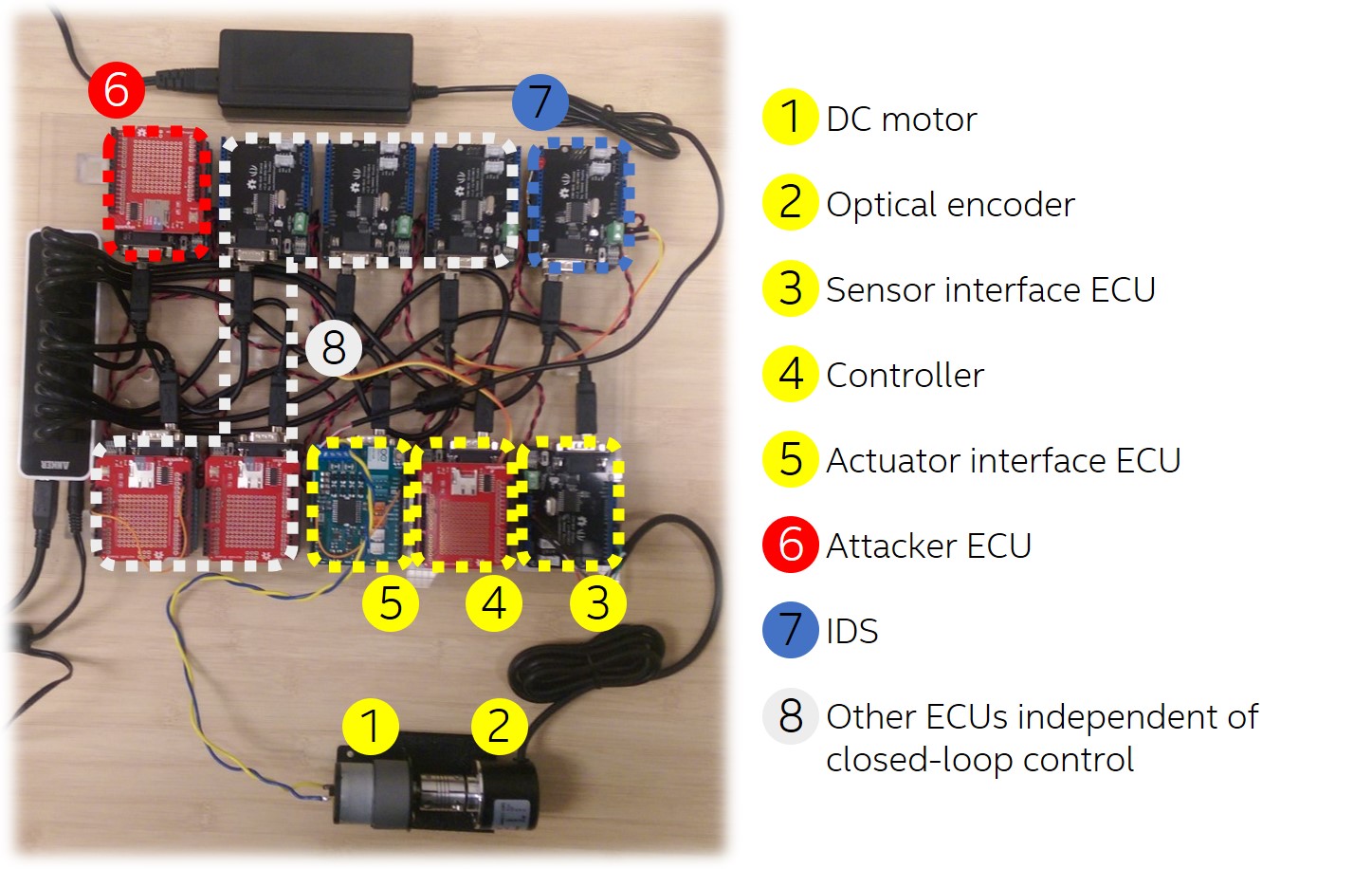}
    \caption{Closed-loop DC motor speed control and intrusion detection and localization on a networked ECU testbed.}
    \label{fig:ETsetup}
    \end{center}
\end{figure}

To illustrate feasibility of the proposed solution, we developed a prototype application that performs closed-loop speed control. The testbed consists of ten Arduino-based ECUs networked through a CAN bus, and corresponding sensing and actuation components. The physical testbed along with the additional components realizing closed-loop DC motor speed control is shown in Figure~\ref{fig:ETsetup}.

\subsection{Architecture and Modeling the ECU Testbed Closed-Loop Control System}
The sensor ECU counts pulses from the optical encoder\footnote{Sensor resolution is $600\:ppr$ (pulses per revolution).} periodically every $50~ms$ (sampling period of the system) when the sensor measurements are sent to the controller ECU. The controller ECU receives this message and consumes the sensor measurements through execution the Proportional-Integral-Derivative (PID) control algorithm computing the control signal. The control signal is a value in the range $[0,2^8-1]$ indicating pulse width of the Pulse Width Modulation (PWM) module to be commanded to the DC motor drive at the actuator ECU. The controller sends this value to the actuator ECU. The actuator ECU receives this message and updates the state of its PWM module internally. This command propagates to the controlled process and is reflected on the speed sensor measurements which are used to close the loop as described.
The plant model (actuation-motor-sensor chain) for purposes of controller and IDS design was obtained through system identification. Speed sensor measurements were acquired for a known PWM input to the actuation stage via serial port directly from ECUs, and were then imported in Matlab where they were fed into the System Identification Toolbox. Obtained first-order model that describes the dynamics of speed in relation to the commanded pulse width is
\begin{equation}
\label{eq:caseStudy}
    x_k^{speed}=0.91x_{k-1}^{DC}+0.095u_{k-1},\\
    y_k^{speed}=x_k^{speed}.
\end{equation}
Since this model features only one state, it will represent the physical state of the speed of the motor; $x_k$ is thus a scalar, rather than a vector. Since the speed is measurable, the output of our process $y_k$ is the same as $x_k$, while $u_k$ is the input (PWM command). A widely used control algorithm (PID controller) has been implemented for speed control; based on this model, PID controller was tuned by using the Simulink PID Tuner in simulation. The PID controller C-code implementation is straightforward and is thus omitted. The parallel implementation of the PID controller was implemented on the controller ECU with parameters \mbox{$K_p=2.5$}, \mbox{$K_i=15.5$}, \mbox{$K_d=0$} (Additional details about PID-based process control can be found in~\cite{PID1,PID2}). Additionally, to avoid wind-up effects due to the integral term of the controller, a simple anti wind-up scheme was implemented as in~\cite{PIDantiWindUp} with the back-calculation gain set to $K_b=0.01$. The controller implementation is not essential for physics-based intrusion detection, but the model of the controller is crucial, and is given by
\begin{align}
    x_k^{ctrl}=&x_{k-1}^{ctrl}+u_{k-1}^{ctrl},\\
    y_k^{ctrl}=&5.775x_k^{ctrl}+3.725u_k^{ctrl}.
\end{align}

Note that the notation for the controller model is unified with the process model, however, $u_k^{ctrl}$ and $u_k$ are not the same signals since the output of the controller $y_k^{ctrl}$ is the actual input of the plant $u_k$. Similarly, the difference between a reference signal and the output of the system $y_k$ is the input of the controller $u_k^{ctrl}$.

Additionally, as described in Section~\ref{subsec:contextAwareness}, both \emph{adaptive thresholding} and \emph{adaptive estimation} are implemented to distinguish disturbances (in this case load on the motor shaft) from attacks. To extract operational context, an additional sensor measuring DC current (for purposes of overload protection) is utilized. Similarly, using the Matlab System Identification Toolbox, the model of the DC current dynamics dependent on the PWM input to the motor is found to be a second order system
\begin{align}
 x_k^{DC}=&\begin{bmatrix}0& 1\\-0.6349& 1.6148\end{bmatrix}x_{k-1}^{DC}+\begin{bmatrix}0.0602\\0.0392\end{bmatrix}u_{k-1},\\
 y_k^{DC}=&\begin{bmatrix}1& 0\end{bmatrix} x_k^{DC}.
\end{align}

Since a single physical system can be described with infinitely many equivalent state-space models, the states of a state-space model do not necessarily have physical meaning. However, by observing the measurement matrix \mbox{$C^{DC}=\begin{bmatrix}1& 0\end{bmatrix}$}, we can conclude that the first state $x_{1,k}^{DC}$ in the state vector $x_k^{DC}=\begin{bmatrix}x_{1,k}^{DC} \\x_{2,k}^{DC}\end{bmatrix}$ is the actual DC current magnitude.

Based on these models, state estimation and residual generation is performed as described in Section~\ref{subsec:kalman}. Kalman filter implementation is straightforward given the two-step estimation equations and is also fully described in~\cite{kalman}. Noise covariance matrices $Q$ and $R$, as appearing in Kalman filter equations, were determined experimentally so that the state estimation gives best observable results. They are experimentally fixed to $Q^{speed}=0.005, R^{speed}=0.1, Q^{ctrl.}=0.5, R^{ctrl.}=0.0005,\\
Q^{DC}=\begin{bmatrix}0.00001& 0\\0& 0.00001\end{bmatrix}, R^{DC}=0.2$,
in the respective order for speed, controller, and DC current estimation. Intuitively, due to high noise in DC current measurements, the model is promoted with lower values in the process noise covariance matrix than the DC current measurements. Similar is true for speed estimation, however, speed measurements are less noisy, which is taken into account accordingly. On the other hand, the controller is precisely implemented in software, and it is guaranteed to conform to the model. Hence, noise covariance values have an opposite trend for the controller model, compared to the plant model.

Three instances of the Kalman filter performing state estimation based on aforementioned models are executed on a separate ECU of the testbed (component 7 in Figure~\ref{fig:ETsetup}), independent of the closed-loop control. Inherently, three residual signals and the DC current measurements are used for attack detection and localization.

\subsection{Experimental Results}
Table~\ref{tab:scenarios} enumerates all experimental scenarios conducted on our case-study testbed. Due to space constraints, signals for an excerpt of scenarios are shown and discussed in the following. In the experiments shown, the attacks are biasing the relevant signals by $+10\%$ of the signals current value.


\begin{table}[]
    \centering
        \caption{Experiment scenario specification}
    \begin{tabular}{p{0.1\linewidth} | p{0.78\linewidth}}
         \hline
         Scenario& Description \\ \hline
         S1& Motor shaft loaded \\
         S2& Speed measurements biased (at sensor ECU) before being transmitted on the bus\\
         S3& Combination of S1 and S2\\
         S4& Biased speed measurements injected into network (by attacker’s ECU)\\
         S5& Combination of S1 and S4\\
         S6& Actuation commands biased before sending to motor (at actuator ECU)\\
         S7& Combination of S1 and S6\\
         S8& Biased actuation commands injected into network (by attacker’s ECU)\\
         S9& Combination of S1 and S8\\
         S10& Sensor measurements biased (at controller ECU) after reception at controller ECU\\
         S11& Combination of S1 and S10\\
         S12& Actuation commands biased (at controller ECU) before sending to actuator ECU\\
         S13& Combination of S1 and S12\\
         \hline
    \end{tabular}
    \label{tab:scenarios}
\end{table}

\begin{figure}[!t]
    \begin{center}
    \includegraphics[width=0.49\textwidth]{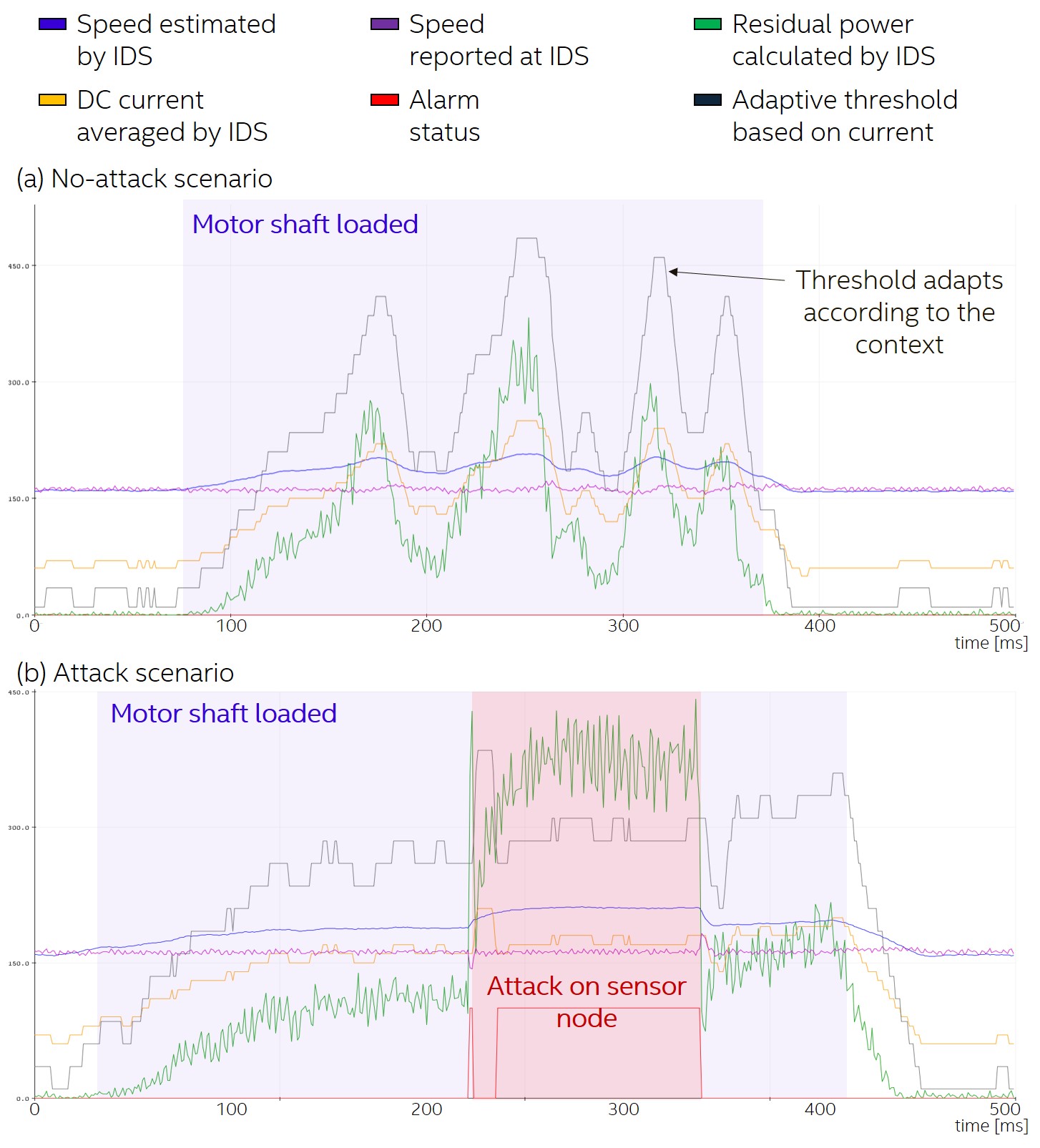}
    \caption{Closed-loop DC motor speed control and intrusion detection using adaptive thresholding under nominal and load conditions, with and without internal sensor attack (signals are scaled to fit the same Y axis for illustration).}
    \label{fig:adaptiveThresholding}
    \end{center}
\end{figure}

We first illustrate effectiveness of our context extraction engine on internal sensor attacks. For adaptive thresholding, the residual threshold value is made linearly dependent on the current value of the DC current consumed by the motor ($g=a\cdot I_{DC}+b$, where $a$ and $b$ are parameters, and $I_{DC}$ is the DC current measurement). Notice that the residual power (green signal) remains below the adaptive threshold (black signal) under load (Figure~\ref{fig:adaptiveThresholding}(a)), but exceeds the threshold under internal sensor attack (Figure~\ref{fig:adaptiveThresholding}(b)).

\begin{figure}[!t]
    \begin{center}
    \includegraphics[width=0.48\textwidth]{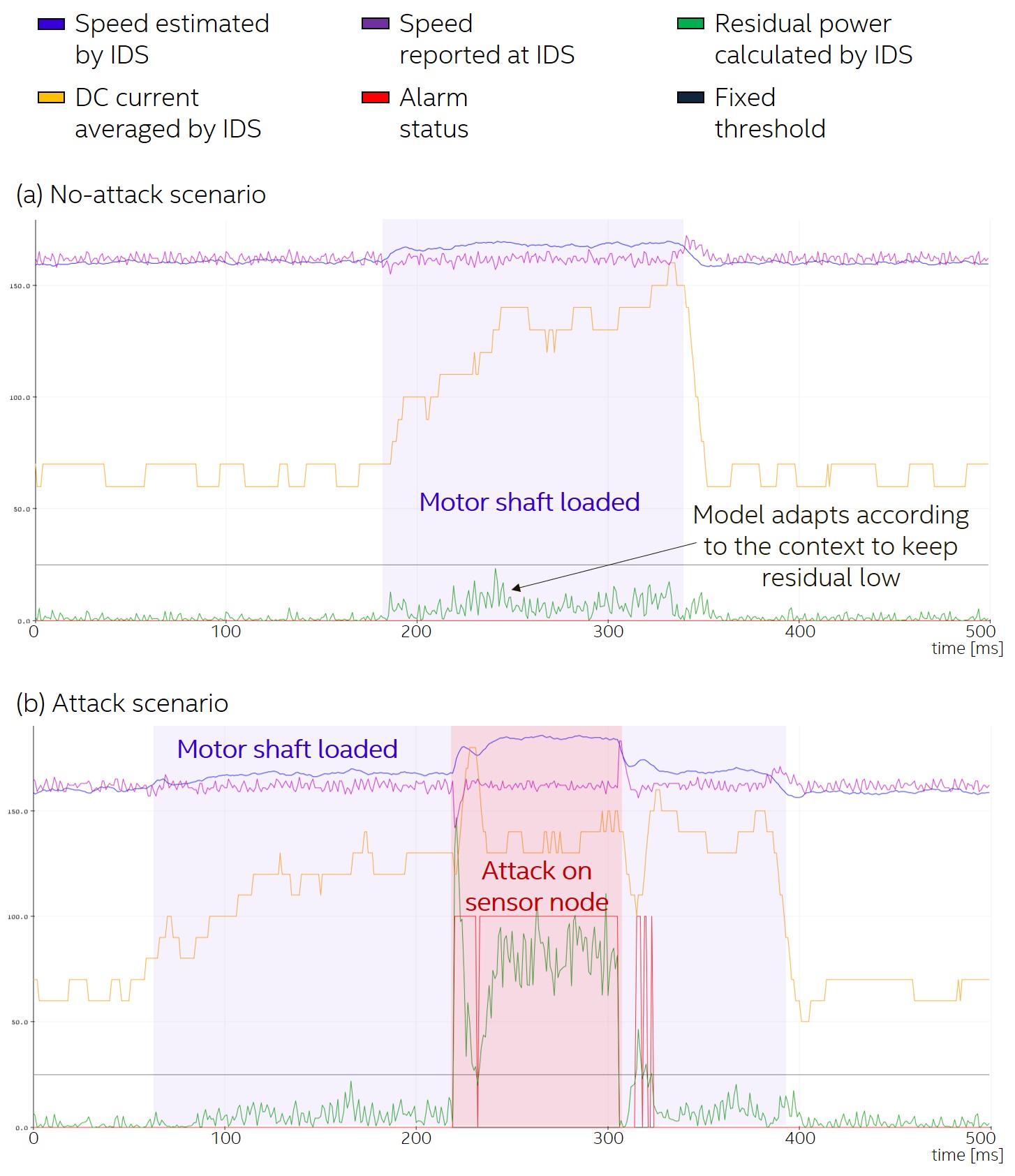}
    \caption{Closed-loop DC motor speed control and intrusion detection using adaptive estimation under nominal and load conditions, with and without internal sensor attack (signals are scaled to fit the same Y axis for illustration).}
    \label{fig:adaptiveEstimation}
    \end{center}
\end{figure}

In the case of adaptive estimation, matrix $B$ of the model~\eqref{eq:caseStudy} is made linearly dependent on the current consumption. Specifically, since $B$ is a scalar for a single input-single output system, a simple relation $B=d\cdot I_{DC}+e$ applies for a range of DC current magnitudes. Notice here that parameter $d$ is negative since, intuitively, this means that as the load increases, the direct proportional effect of the input voltage on the speed of the motor decreases. Recall that matrix $B$ in the model~\eqref{eq:LTI} captures how the system’s state (motor speed in this case) depends on the input (motor’s pulse width control signal). Figure~\ref{fig:adaptiveEstimation}(a) shows how the residual signal remains below the fixed threshold even under disturbance conditions, but exceeds the threshold under internal sensor attack in Figure~\ref{fig:adaptiveEstimation}(b).

\begin{figure}[!t]
    \begin{center}
    \includegraphics[width=0.49\textwidth]{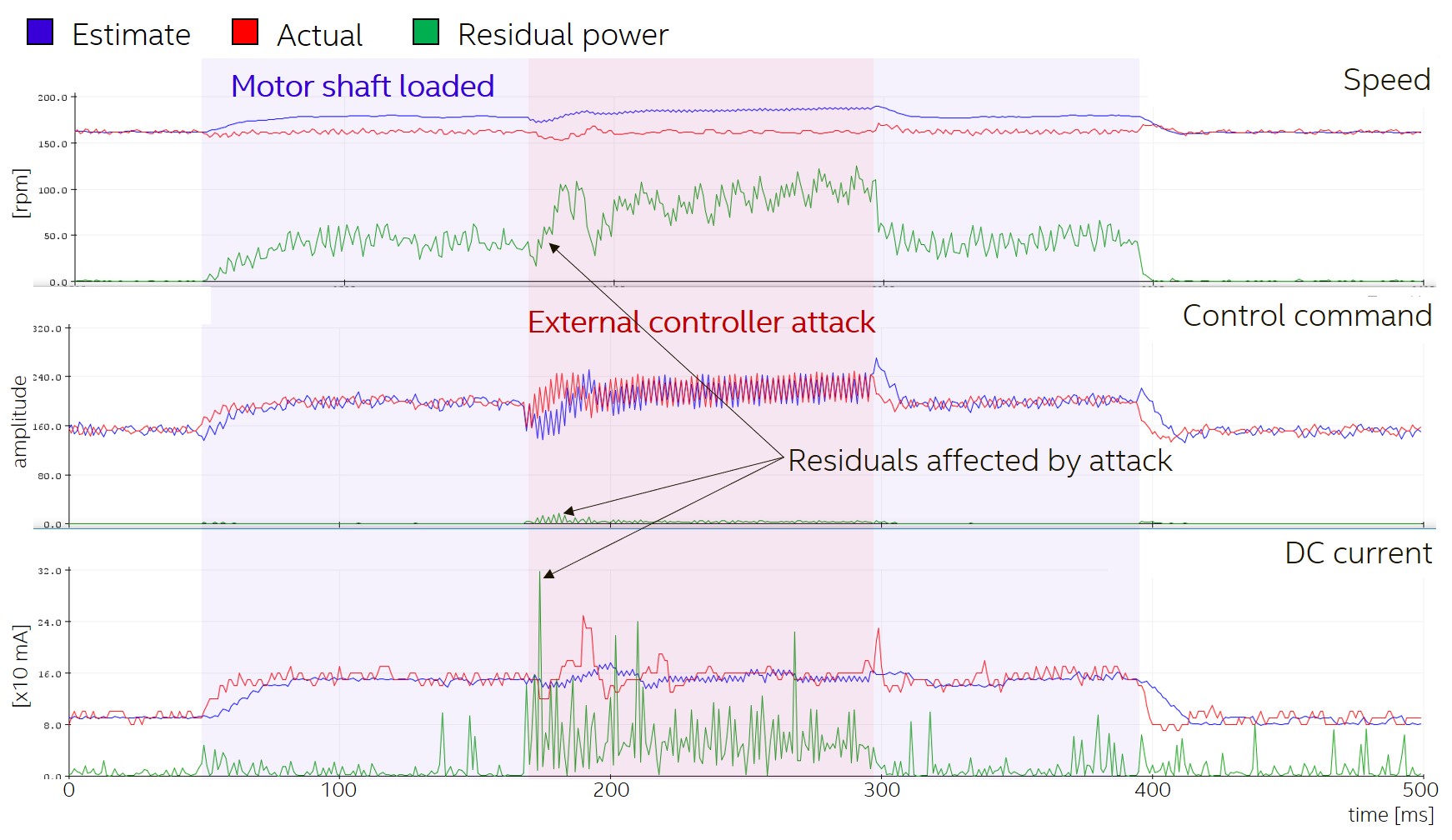}
    \caption{Closed-loop DC motor speed control and intrusion detection under nominal and load conditions, with and without external controller attack.}
    \label{fig:cs_externalControllerAttack}
    \end{center}
\end{figure}

Furthermore, Figure~\ref{fig:cs_externalControllerAttack} shows detector performance under external controller attack (similarly under load). Notice that now the control command (i.e., controller) residual is affected, in addition to the speed (i.e., plant) residual when the attack is active. Ultimately, we are able to deduct a mapping between the pattern of affected signals within the IDS, and the attack target and class.

\subsection{Attack Characterization}
Table~\ref{tab:localization} shows localization success results for attacks from all scenarios defined in Table~\ref{tab:scenarios}. At run-time, the pattern of affected signals can be matched to the corresponding entry in Table~\ref{tab:scenarios}, unveiling the attack target and class. If state-of-the art physics-based intrusion detection was implemented, firstly, the attack model would have to be reduced to assume the controller is secure. Secondly (and consequently), only internal sensor and actuator attacks could be detected (but not distinguished). Additionally, a fixed threshold over the Kalman-based residuals would raise false alarms in cases when operational context is different from nominal (e.g., motor under load). On the other hand, our approach is able to not only detect more types of attacks, but is also able to distinguish them; this is a critical requirement towards steps following attack detection. While we discuss the graphical features of residual patterns, other approaches (e.g., based on machine learning) can easily be employed.

\begin{table}[]
    \centering
        \caption{Attack characterization mapping for experimental scenarios.}
    \begin{tabular}{p{0.08\linewidth} | p{0.07\linewidth} | p{0.09\linewidth} | p{0.05\linewidth} | p{0.45\linewidth}}
             \multicolumn{4}{c|}{Impacted parameters}& \\
             $r_{speed}$& $r_{ctrl}$& DC current& $r_{DC}$& Attack \\ \hline
         $\times$& & & & Internal sensor or internal actuator attack w/o load\\
         $\times$& &$\times$ & & Internal sensor or internal actuator attack w/ load\\\hline
         & $\times$ & &$\times$ & External sensor attack w/o load\\
         & $\times$ &$\times$ &$\times$ & External sensor attack w/ load\\\hline
         & $\times$& & & Internal controller attack w/o load\\
        $\times$& $\times$& $\times$& & Internal controller attack w/ load\\\hline
        $\times$& $\times$& &$\times$ & External controller attack w/o load\\
        $\times$& $\times$&$\times$ &$\times$ & External controller attack w/ load\\
    \end{tabular}
    \label{tab:localization}
\end{table}

\section{Related Work and Discussion}
\label{sec:relatedWork}

Multiple types of intrusion detectors targeting in-vehicle networks have been proposed. For instance, frequency-based IDSs deal with a specific type of attack where additional messages are injected by the attacker (e.g.,~\cite{gutierrez_DSN19}). Package injection disturbs the normal pattern of periodic transmissions that are already scheduled in the network and that is used as a principle to detect intrusion. Though that technique can address attacks against periodic messages, it is not effective for aperiodic ones. Other solutions take into account more features of the bus traffic, such as inter-message correlations. It relies on a training phase that is assumed to occur during a non-attack condition, to result in legitimate bus pattern models. Re-training sessions may be required to compensate for different operational contexts. Detection scope is limited by the richness of traffic features identified during the (re-)learning phase. This directly affects the false positive rate.

Other schemes like ECU voltage fingerprinting focus on the authenticity of the transmitter (e.g.,~\cite{ahmed_ESCAR19}). Though this can protect against ECU impersonation, it cannot detect attacks that originate from software execution on a legitimate module. More specifically, it does not inspect message contents, which makes it impossible to determine whether a legitimate ECU is sending malicious data. Moreover, the main drawback of approaches based on machine learning techniques is that they are usually computationally demanding which consequently introduce detection delay. Additionally, presence of an attack can be flagged by these techniques, however localization of the attack is limited. 

Cryptographic methods such as signatures and Message Authentication Codes (MACs) by definition require processing time and additional bus bandwidth to transmit the authentication tags. Cryptographic mechanisms require key management schemes, which represents an additional challenge. These requirements are not supported by legacy ECUs and networks (such as CAN). Additionally, these techniques only act preventively for masquerading attacks originating from an untrusted entity who has network access and are not effective for attackers who gained access to software execution on an existing trusted ECU.

A body of existing work also explores enhancing security in legacy networks (e.g., based on CAN bus) by retrofitting of authentication services (e.g.,~\cite{lesi_TCPS20}) or modifying the execution paradigm (e.g.,~\cite{scheduleObfurscation}). All these works rely on the ability to perform some level of modification to the firmware executed by legacy ECUs (e.g., running additional security services, performing changes to real-time scheduling policies, etc.), which requires prohibitive changes to implementations that are commonly firmly defined. On the other hand, our centralized IDS can be implemented on a trusted platform and ‘plugged-in’ to an existing non-trusted network. Additionally, no fine-grained implementation details of existing ECUs are required, as these are not practically available except to the ECU manufacturer.

In the domain of physics-based intrusion detection, it is worth noting that residual signals have been used in the past for failure localization. They can be carefully computed based on perturbed models that already include a specific failure; this way anomaly localization can also be performed based on the exact residual signal that exceeds a threshold~\cite{FDI3,FDI4}. However, knowledge of the physical models under different anomalous conditions is needed, and the assumption that an anomaly will follow a specific dynamical behavior can be valid for failures, but not for attacks.

Detection limitations for our approach arise from two major sources. Firstly, zero dynamics~\cite{zeroDynamics} is the intrinsic dynamics of the system that limits the outputs to zero (i.e., it is not detectable by sensors), while system states are indeed changing; this work does not consider zero dynamics attacks. Secondly, noises in realistic systems allow the attacker to introduce \emph{slow} attacks without obviously perturbing the system more than what is considered normal operation, while accumulating significant error in the system’s trajectory over longer periods of time. This \emph{stealthy} adversary must have knowledge of the physical process and the deployed control algorithms, \emph{as well as} the implemented physics-based IDS and all the utilized models. This knowledge enables an attacker to inject false signals or alter existing signals in such a manner that residual signals do not exceed thresholds. However, depending on the attacker's objectives and the dynamics of the system at hand, it may be impossible to introduce significant attacks during a reasonable amount of time.
\section{Conclusion}
\label{sec:conclusion}
In this paper, we extended physics-based intrusion detection by (1) considering a more comprehensive attack model than previous literature, (2) enabling a network-attached intrusion detection implementation that, unlike existing physics-based approaches, does not rely on security of legacy nodes, and (3) providing attack detection and localization. We achieved this by utilizing controller state estimation (in addition to monitoring plant's states) and residual signal patterning. Specifically, by matching the pattern of affected signals with a predefined table, we are able to characterize the target and class of attack. Additionally, by utilizing other sensors outside of the monitored control loop, we developed a context-aware intrusion detector that distinguishes false alarms due to operational conditions from attacks by utilizing either adaptive thresholding or adaptive estimation. We validated our proposal both in simulation, on an automated lane keeping system, and experimentally, on a DC motor speed control application running on top of a fully distributed, CAN bus-based ECU testbed. Though it has been observed that automotive systems have shown some fragility against cyber threats, the capability of detecting and characterizing attacks in autonomous systems is an important feature to increase security, and ultimately enhance overall safety.


As avenue for future work, we may extend our considerations to zero dynamics attacks, as well as consider how wear and tear during the system's lifetime affects the correctness of physical models, and consequently performance of intrusion detection. To satisfy the constant need for a correct model, on-line closed-loop system identification and/or model parameter estimation can be applied. However, in addition to functional aspects of these techniques, secure model identification/parameter estimation is another challenge since estimating parameters based on potentially spoofed data is equivalent to having an inaccurate model.

\bibliographystyle{IEEEtran}
\bibliography{refs}

\begin{thebibliography}{10}
\bibitem{jeepHack}
------, ``The full story of how that {Jeep} was hacked.''
\end{thebibliography}

\end{document}